\documentclass[final,11pt,twocolumn,sort&compress]{elsarticle} 

\usepackage{listings}
\usepackage{amsmath}
\usepackage{graphicx}
\usepackage{color}
\usepackage{setspace}
\usepackage{url}
\usepackage{widetext} 
\usepackage{alltt, parskip, fancyhdr, boxedminipage}
\usepackage{makeidx, multirow, longtable, tocbibind, amssymb}
\usepackage{hyperref} \hypersetup{colorlinks=true,citecolor=blue,linkcolor=blue,urlcolor=blue} 
\usepackage{enumitem}

\usepackage{relsize} 
\newcommand{\AFLOWpi}{ {\sf AFLOW$\mathlarger{\mathlarger{\mathlarger{{\pi}}}}$}}
\newcommand{\PAOFLOW}{ {\sf PAOFLOW}}

\newcommand{\V}[1]{\vec{#1}}

\newcommand{\bk}{\mathbf{k}}

\newlength{\funcindent}
\newlength{\funcwidth}
\DeclareFixedFont{\ttb}{T1}{txtt}{bx}{n}{12} 
\DeclareFixedFont{\ttm}{T1}{txtt}{m}{n}{12}  

\usepackage{color}
\definecolor{deepblue}{rgb}{0,0,0.5}
\definecolor{deepred}{rgb}{0.6,0,0}
\definecolor{deepgreen}{rgb}{0,0.5,0}

\def\AFLOW{{\small AFLOW}}   

\newcommand\pythonstyle{\lstset{
basicstyle=\ttb\scriptsize,	
keywordstyle=\ttb\scriptsize\color{deepblue},
commentstyle=\ttb\scriptsize\color{deepred},    
stringstyle=\ttb\scriptsize\color{deepgreen},
 language=Python,
  aboveskip=0mm,
  belowskip=0mm,
  showstringspaces=false,
  columns=flexible,
  numbers=none,
  numberstyle=\tiny\color{gray},
  ndkeywords={return, class, if ,elif, endif, while, do, else, True, False , catch, def},
  ndkeywordstyle=\color{red}\bfseries,
  breaklines=true,
  postbreak=\raisebox{0ex}[0ex][0ex]{\ensuremath{\color{black}\hookrightarrow\space}},
  breakatwhitespace=True,
  tabsize=2
}}

\lstnewenvironment{python}[1][]
{
\pythonstyle
\lstset{#1}
}
{}

\newcommand\pythoninline[1]{{\pythonstyle\lstinline!#1!}}

\newcommand\fortranstyle{\lstset{
language=Fortran,
basicstyle=\ttb\scriptsize,
otherkeywords={self},             
keywordstyle=\ttb\scriptsize\color{deepblue},
emph={MyClass,__init__},          
commentstyle=\ttb\scriptsize\color{deepred},    
stringstyle=\ttb\scriptsize\color{deepgreen},
showstringspaces=false,
}}

\lstnewenvironment{fortran}[1][]
{
\fortranstyle
\lstset{#1}
}
{}

\newcommand\fortraninline[1]{{\fortranstyle\lstinline!#1!}}

\usepackage{listings}
\usepackage[most]{tcolorbox}
\usepackage{inconsolata}

\newtcblisting[auto counter]{sexylisting}[2][]{sharp corners, 
    fonttitle=\bfseries, colframe=gray, listing only, 
    listing options={basicstyle=\fontsize{9}{11}\ttfamily,language=Python}, 
    title=Listing \thetcbcounter: #2, #1}

\definecolor{Gray}{gray}{0.9}
\definecolor{LightCyan}{rgb}{0.88,1,1}
\begin{document} 
\begin{frontmatter} 



\title{{\bf \LARGE Advanced modeling of materials with {\sf PAOFLOW} 2.0:
\\New features and software design}}

\author{Frank T. Cerasoli$^1$, Andrew R. Supka$^{2,3}$, Anooja Jayaraj$^1$, Marcio Costa$^4$, Ilaria Siloi$^{5}$, Jagoda S\l awi\'{n}ska$^6$, Stefano Curtarolo$^{7}$, Marco Fornari$^{2,3,7}$ Davide Ceresoli$^{8}$, and Marco Buongiorno Nardelli$^{1,7,\star}$}
\address{$^{1}$ Department of Physics and Department of Chemistry, University of North Texas, Denton TX, USA }
\address{$^{2}$ Department of Physics, Central Michigan University, Mount Pleasant MI, USA}
\address{$^{3}$ Science of Advanced Materials Program, Central Michigan University, Mount Pleasant MI, USA}
\address{$^{4}$ Department of Physics, Fluminense Federal University, Niterói 24210-346, Rio de Janeiro, Brazil}
\address{$^{5}$ Department of Physics and Astronomy, University of Southern California, Los Angeles, CA 90089, United States}
\address{$^{6}$ Zernike Institute for Advanced Materials, University of Groningen, Nijenborgh 4, NL-9747AG Groningen, NL}
\address{$^{7}$ Center for Materials Genomics, Duke University, Durham, NC 27708, USA}
\address{$^{8}$Consiglio Nazionale Delle Ricerche, Istituto di Scienze e Tecnologie Chimiche "G. Natta" (CNR-SCITEC), 20133 Milan, Italy}
\address{$^{\star}${\bf corresponding:} mbn@unt.edu}

\begin{abstract}
Recent research in materials science opens exciting perspectives to design novel quantum materials and devices, but it calls for quantitative predictions of properties which are not accessible in standard first principles packages. \PAOFLOW\, is a software tool that constructs tight-binding Hamiltonians from self-consistent electronic wavefunctions by projecting onto a set of atomic orbitals. The electronic structure provides numerous materials properties that otherwise would have to be calculated via phenomenological models. In this paper, we describe recent re-design of the code as well as the new features and improvements in performance. In particular, we have implemented symmetry operations for unfolding equivalent {\bf k}-points, which drastically reduces the runtime requirements of first principles calculations, and we have provided internal routines of projections onto atomic orbitals enabling generation of real space atomic orbitals. Moreover, we have included models for non-constant relaxation time in electronic transport calculations, doubling the real space dimensions of the Hamiltonian as well as the construction of Hamiltonians directly from analytical models. Importantly, \PAOFLOW\ has been now converted into a Python package, and is streamlined for use directly within other Python codes. The new object oriented design treats \PAOFLOW's computational routines as class methods, providing an API for explicit control of each calculation.
\end{abstract}


 \end{frontmatter} 

\section{Introduction}
\label{introduction}
Exploring phenomena and properties of novel materials requires accurate and efficient computational tools that can be easily customized and manipulated. In this context, {\it ab initio} tight-binding (TB) Hamiltonians constructed from self-consistent quantum-mechanical wavefunctions projected onto a set of atomic orbitals have been very successful, since they allow calculations for materials that cannot be properly addressed using only density functional theory (DFT) such as large moiré superstructures or properties of exotic quantum systems where spin and topology play an important role. \PAOFLOW~ \cite{PAOFLOW:2018}~is a new software tool that employs an efficient procedure of projecting the full plane-wave solution on a reduced space of pseudoatomic orbitals \cite{Agapito:2016jl, Agapito:2016en}, and provides an interpolated electronic structure to promptly compute a plethora of relevant quantities, including optical and magnetic properties, charge and spin transport as well as topological invariants. Importantly, in contrast with other common approaches the projection does not require any additional inputs and can be successively integrated in high-throughput calculations of arbitrary complex materials. The code has been employed in multiple areas of materials science since its initial release in 2016. In particular, several groups used it to compute the (spin) Berry curvature as well as spin and/or anomalous Hall conductivity (SHC and AHC) in a variety of materials, ranging from  $\beta$-W to magnetic antiperovskites \cite{PhysRevMaterials.4.094404, huyen2021spin, PhysRevLett.122.077203, Nan2020, PhysRevMaterials.3.044409}. Transport quantities, such as the electrical and thermal conductivity, were also computed in order to analyze carrier mobility in thermoelectrics \cite{doi:10.1021/acs.chemmater.1c00872, D1TA01615F}.

The software package has recently undergone a major refactor, resulting in a large variety of properties that can be calculated as well as highly improved performance. Many improvements were made to simplify the user experience, to make the package more modular, and to create an API for manipulating TB Hamiltonians.\PAOFLOW, now installed as a Python package, features an object oriented design and contains an importable\PAOFLOW\ class, allowing multiple Hamiltonians to be constructed and manipulated simultaneously. This framework enables high-throughput materials analysis within a single python file. While explicit descriptions of the code's methodology are available in the original\PAOFLOW\ paper \cite{PAOFLOW:2018}, in this paper we outline\PAOFLOW's modified features, detail new functionalities, and provide a user manual for operating the various methods available within the package. Currently, \PAOFLOW\ is publicly available under the terms of the GNU General Public License as published by the Free Software Foundation, either version 3 of the License, or any later version. It is also integrated in the \AFLOWpi\ high-throughput framework \cite{Supka:2017kd} and distributed at \verb|http://www.aflow.org/src/aflowpi| and \verb|http://www.aflow.org/src/paoflow|  \cite{Curtarolo:2012fa, curtarolo:art75}.

\section{Installation and software design}
\label{design}
\PAOFLOW\ is written in Python 3.8 (using the Python standard libraries, \verb|NumPy| and \verb|SciPy|).
Parallelization on CPUs uses the openMPI protocol through the \verb|mpi4py| module.
The \PAOFLOW\ package can be easily installed on any hardware. Installation directly from the Python package index (PyPi) is possible with the single command: {\tt pip install paoflow}. Otherwise, one may clone the\PAOFLOW\ repository from GitHub and install it from the root directory with the command: {\tt python setup.py install}. The package requires no specific setup, provided that the prerequisite Python 3.8 modules are installed on the system. Example codes illustrating\PAOFLOW's capabilities are included on GitHub, in the {\tt examples/} directory of the\PAOFLOW\ repository. Once installed,\PAOFLOW\ can be imported into any Python code and used in conjunction with other software packages. It should be noted that, in \PAOFLOW\ 1.0 two control files were required: {\tt main.py} to begin the execution, and {\tt inputfile.xml} to provide details of the calculation. In\PAOFLOW\ 2.0 the desired routines are called directly from the Python code, eliminating the need for {\tt inputfile.xml}. To facilitate the transition between the versions, in the {\tt examples/} directory we have provided an updated {\tt main.py} file which allows use of the version 1.0 XML inputfile structure in\PAOFLOW\ 2.0.\\

Generally,\PAOFLOW\ requires a few basic calculations performed with the Quantum ESPRESSO (QE) package \cite{Giannozzi:2017io}. The first (self-consistent) run generates a converged electronic density and Kohn-Sham (KS) potential on an appropriate Monkhorst and Pack (MP) {\bf k}-point mesh (\verb|pw.x|). The second (non self-consistent) one (\verb|pw.x|) evaluates eigenvalues and eigenfunctions on a larger MP mesh and often for an increased number of bands. In \PAOFLOW\ 1.0 the latter required full {\bf k}-point mesh without using symmetries, i.e. setting flags {\tt nosym=.true.} and {\tt noinv=.true.} in QE inputfile, while version 2.0 can reconstruct equivalent {\bf k}-points from symmetry operations and has no such requirement, which greatly reduces the computation time at the level of DFT. As a next step, the KS wavefunctions from QE must be projected onto PAO basis functions. Also here we have introduced an important change - a new capability is the generation of real space atomic orbitals, constructed from the product of radial components from pseudopotential files and spherical harmonics specifying angular momentum dependence. The user can thus choose between two different options: project the KS wavefunctions onto this internally constructed PAO basis set (Sec. \ref{projections}), or post-processing the QE output with \verb|projwfc.x| in order to perform the projection, as it was done in version 1.0. If the projections are computed using QE, the eigenfunctions must be read by \PAOFLOW\ before the construction of a PAO Hamiltonian (Sec. \ref{read_atomic_proj_QE}). Finally, \PAOFLOW\ 2.0 introduces a possibility to operate with no preprocessing requirements from QE, where built-in or user-defined models serve as the recipe for building Hamiltonians. Implementing these models requires specific information about the atomic system {\it a priori}, and their usage is described in Section \ref{tbmodels}.

Central to \PAOFLOW\ 2.0 is an internal object called the {\tt DataController}, which has the sole responsibility of collecting and maintaining important information about the atomic system and its corresponding Hamiltonian. The {\tt DataController} is initially populated with data from the QE calculation, providing many of \PAOFLOW's routines with required information and simplifying the way of calling functions by the user. Moreover, the {\tt DataController} saves quantities needed for subsequent calculations, such as the Hamiltonian's gradient or the adaptive smearing parameters. In fact, many routines require calculation of some quantities as a prerequisite. For example, the {\tt spin\_Hall} routine requires the Hamiltonian's gradient, which means that the {\tt gradient\_and\_momenta} should be called first to populate the {\tt DataController} with the gradient and momenta. The {\tt DataController} stores the system information in two dictionaries: one for strings and scalar attributes ({\tt data\_attributes}) and another for vector and tensor quantities ({\tt data\_arrays}). Such a structure allows computed quantities to be easily accessed from \PAOFLOW\ and utilized in customized calculations defined by the user. Note that the dictionary keys are consistent with the naming conventions of \PAOFLOW\ 1.0 to facilitate backwards compatibility with the XML inputfiles and minimize differences in the user experience when transitioning from the previous version. 

\section{Code description and package usage}
\label{code}

\PAOFLOW's most fundamental procedure is the construction of accurate PAO Hamiltonians, and the code's object-oriented design allows users to manipulate multiple Hamiltonians easily. A \PAOFLOW\ object is responsible for a single Hamiltonian, which is constructed and operated on with \PAOFLOW's class methods. If instead of DFT electronic wavefunctions a TB model is used to construct the Hamiltonian, the new \PAOFLOW\ object will create the Hamiltonian immediately. Otherwise, the KS wavefunctions are read from the output of QE. The atomic orbitals are constructed and the KS wavefunctions are projected onto them with the {\tt projections} routine, creating the PAO basis. If the projections are performed by QE's module {\tt projwfc.x}, they must be read explicitly with {\tt read\_atomic\_proj\_QE}. Then, the Hamiltonian is constructed with {\tt build\_pao\_hamiltonian}, which allows \PAOFLOW's other class methods to become functional. Listing 1 provides an example source code for building the \PAOFLOW\ object, reading projections performed by QE, and constructing the PAO Hamiltonian. Listing 2 performs the same initialization procedure, but uses the internal atomic orbital projection scheme. An ellipsis appearing in any listing indicates that other \PAOFLOW\ routines may follow.

The following subsections outline \PAOFLOW's individual routines and the arguments that they accept for control. These routines belong to the file PAOFLOW.py, located in the package's {\tt src/} directory, and should be called directly by the user. 

\subsection{The constructor: {\tt PAOFLOW}}
\label{importandconstructor}

The\PAOFLOW\ constructor acquires information about the python execution, the names of input/output/working directories, and about the atomic system. It builds and populates the {\tt DataController}, which will maintain the important quantities involved in calculations, handle communication in multi-core runs, and write files to disc when necessary. 

For\PAOFLOW\ to build a Hamiltonian, the constructor must be passed data with either the location of Quantum ESPRESSO's {\tt .save} directory or required specifications for an analytical TB model. The QE {\tt .save} directory contain up to two files from the execution of QE: {\tt data-file-schema.xml} ({\tt data-file.xml} in previous versions of QE) generated by the main run with {\tt pw.x}, and {\tt atomic\_proj.xml} generated by the post-processing tool {\tt projwfc.x}, if the projections are computed with QE. To implement a TB model from scratch, rather than starting from the KS wave function solutions of DFT, a dictionary containing the model's label and other required parameters should be passed into the {\bf model} argument (see Section \ref{tbmodels}).
\\\\
Arguments for the constructor, {\tt PAOFLOW}:
\begin{itemize} 
\item {\bf workpath} (string) - {\it Default}: {\tt './'} - Path to the working directory. Defaults to the current working directory.
\item {\bf outputdir} (string) - {\it Default}: {\tt 'output'} - Name of the directory to store output data files. The directory is created automatically in the {\bf workpath}, if it does not already exist. 
\item {\bf inputfile} (string) - {\it Default}: {\tt None} - This argument is primarily for backwards compatibility with\PAOFLOW\ 1.0. It names the XML inputfile with control parameters described in Ref. \cite{PAOFLOW:2018}. The XML inputfile provides also a consistent descriptor format for highly automated calculations, utilized by AFLOW$\pi$.
\item {\bf savedir} (string) - {\it Default}: {\tt None} - Name of the Quantum ESPRESSO {\tt .save} directory, relative to the working directory.
\item {\bf model} (dict) - {\it Default}: {\tt None} - Dictionary specifying parameters required to implement a TB model. See Section \ref{tbmodels}.
\item {\bf npool} (integer) - {\it Default}: {\tt 1} - Number of batches to process when communicating between processors. This value will be automatically increased if the Hamiltonian size exceeds {\tt mpi4py}'s limit for a single cross core message.
\item {\bf smearing} (string) - {\it Default}: {\tt 'gauss'} - Selects the broadening technique used to smooth computed quantities. Options include {\tt 'gauss'}, {\tt 'm-p'} (Methfessel-Paxton), and {\tt None}.
\item {\bf acbn0} (bool) - {\it Default}: {\tt False} - Read overlaps to construct a non-orthogonal PAO Hamiltonian. Necessary to perform ACBN0 calculations \cite{Agapito:2015iz, Gopal:2015bf}.
\item {\bf verbose} (bool) - {\it Default}: {\tt False} - Flag for high verbosity. Set {\tt True} to include additional information in the\PAOFLOW\ output.
\item {\bf restart} (bool) - {\it Default}: {\tt False} - Indicates the continuation of a previous run from the saved state. Once the\PAOFLOW\ object is instantiated, the {\tt restart\_load} routine should be called with the save file's prefix as an argument. An example is provided in listings 4 and 5.
\end{itemize}

\subsection{\tt projections}
\label{projections}
Perform projections of the KS eigenfunctions onto the pseudoatomic orbital basis which is constructed by \PAOFLOW\ based on the information in the atomic pseudopotential files. This operation requires that \PAOFLOW\ is instantiated by passing a QE {\tt .save} directory with required XML file from the self-consistent and non self-consistent calculations. Listing 2 provides example usage of the {\tt projections} routine, and a complete description of the projection methodology can be found in the Appendix of Ref. \cite{Agapito:2016jl}.

\subsection{\tt read\_atomic\_proj\_QE}
\label{read_atomic_proj_QE}
Read the projections of KS wavefunctions onto the atomic orbital basis of the pseudopotential, performed by the QE routine {\tt projwfc.x} and written to {\tt atomic\_proj.xml}. Any time that the Hamiltonian is built directly from the projections of QE, this routine should be called immediately after the constructor.

{\it read\_atomic\_proj\_QE} does not accept any arguments.

\subsection{\tt projectability}
The {\tt projectability} routine determines which bands do not meet projectability requirements, flagging them for shift into the null space. The projectability $p_{\bf k}$ is a quantity measuring how well a KS Bloch state is represented by orbitals of the PAO basis, as described in Section 3 of Ref. \cite{PAOFLOW:2018}. If $p_{\bf k}\approx 1$, the PAO basis set accurately represents that particular Bloch state for ${\bf k}$. $p_{\bf k}\ll 1$ indicates that the state is poorly represented and should be excluded. The projection threshold {\bf pthr} selects the minimum allowed projectability for accepting a band. All bands which do not meet the criteria are projected to the null space during the Hamiltonian's construction, in one of two ways. Either as
\begin{equation}
\hat{H}(\bk) = AEA^\dagger + \kappa \left( I-A A^{\dagger}\right)
\label{eq:Hk13}
\end{equation}
following Ref. \cite{Agapito_2013_projectionsPRB} or
\begin{equation}
\hat{H}(\bk) = AEA^\dagger + \kappa \left( I-A \left( A^{\dagger}A \right)^{-1}A^\dagger \right)
\label{eq:Hk16}
\end{equation}
according to Ref. \cite{Agapito:2016jl}. Unless the {\bf shift} argument is explicitly set to a floating point value, the shifting parameter $\kappa$ is determined automatically by this routine. Which method is used to remove low-projectability bands during the Hamiltonian construction is selected by argument {\bf shift\_type}, in the {\tt pao\_hamiltonian} routine (Sec. \ref{pao_hamiltonian}).

Arguments for {\tt projectability}:
\begin{itemize} 
\item {\bf pthr} (float) - {\it Default}: {\tt 0.95} - The projectability threshold. All bands with a minimum projectability of the {\bf pthr} value or higher are included in the Hamiltonian.
\item {\bf shift} (string or float) - {\it Default}: '{\tt auto}' - Float to indicate the value (in eV) of the null space cutoff ($\kappa$ in eqs. \ref{eq:Hk13} and \ref{eq:Hk16}). Bands beneath the projectability threshold will be shifted to this value. Providing the default argument '{\tt auto}' automatically sets {\bf shift}'s value to the minimum energy of the first band that fails the projectability threshold.
\end{itemize}

\subsection{\tt pao\_hamiltonian}
\label{pao_hamiltonian}
This routine constructs the Hamiltonian in both real space and momentum space. After this routine is completed, the data controller will contain arrays {\tt HRs} and {\tt Hks}, for the respective real space and {\bf k}-space Hamiltonians.\\
\begin{sexylisting}{{\tt main.py} - Build Hamiltonian}
from PAOFLOW import PAOFLOW

pao = PAOFLOW.PAOFLOW(savedir='system.save')
pao.read_atomic_proj_QE()
pao.projectability(pthr=0.95)
pao.pao_hamiltonian()
...
\end{sexylisting}\\
Arguments for {\tt pao\_hamiltonian}:
\begin{itemize} 
\item {\bf shift\_type} (integer) - {\it Default}: {\tt 1} - Determines which method (Eq. \ref{eq:Hk16} by default) is used to remove bands into the null space.\\ 0 - Eq. \ref{eq:Hk13}, 1 - Eq. \ref{eq:Hk16}, or 2 - No shift.
\item {\bf insulator} (bool) - {\it Default}: {\tt False} - Setting this flag as {\tt True} asserts that the system is insulating, setting the top of the highest occupied band to 0 eV. The Fermi energy is calculated for metallic systems, which corrects numerical discrepancies from the projection routine and irreducible wedge unfolding. This flag is set {\tt True} automatically if the QE ouput does not contain smearing parameters.
\item {\bf write\_binary} (bool) - {\it Default}: {\tt False} - Flag to write the files necessary for the ACBN0 routine \cite{PhysRevX.5.011006}. Overlaps from {\tt projwfc.x} are required from prerequisite QE calculations, and this flag is required for further use with ACBN0.
\item {\bf expand\_wedge} (bool) - {\it Default}: {\tt True} - Applies symmetry operations to the {\bf k}-mesh unfolding the irreducible wedge into the complete Brillouin zone, and evaluates the Hamiltonian matrix elements for every {\bf k}-point. \PAOFLOW\ routines act on the full grid of {\bf k}-points ({\tt True}), while ACBN0 only requires the irreducible wedge ({\tt False}).
\item {\bf symmetrize} (bool) - {\it Default}: {\tt False} - The Hamiltonian incurs numerical errors during the process of unfolding the wedge. Certain routines, such as {\tt find\_weyl\_points}, are sensitive to the Hamiltonian's symmetric components. Setting this flag to {\tt True} symmetrizes the Hamiltonian with an iterative procedure to reduce numerical errors in the Hamiltonian's symmetry.
\item {\bf thresh} (float) - {\it Default}: {\tt 1e-6} - The tolerance of symmetrization, if the procedure is performed.
\item {\bf max\_iter} (integer) - {\it Default}: {\tt 16} - The maximum number of iterations that the symmetrization procedure will perform.
\end{itemize}

\subsection{\tt bands}
Computes the band structure along the \AFLOW\ standard path for the specified Bravais lattice. Alternatively, a custom path can be created by defining the high symmetry points in a dictionary and the band path as a string (see Listing 2). The path is Fourier interpolated to an arbitrary resolution, controlled by argument {\bf nk}.
\begin{sexylisting}{{\tt main.py} - Bands}
from PAOFLOW import PAOFLOW

pao = PAOFLOW.PAOFLOW(savedir='system.save')
pao.projections()
pao.projectability()
pao.pao_hamiltonian()

path = 'G-X-S-Y-G'
sym_points = {'G':[0.0, 0.0, 0.0],
              'S':[0.5, 0.5, 0.0],
              'X':[0.5, 0.0, 0.0],
              'Y':[0.0, 0.5, 0.0]}
pao.bands(ibrav=8,
          nk=1000,
          band_path=path, 
          high_sym_points=sym_points)
...
\end{sexylisting}\\
Arguments for {\tt bands}:
\begin{itemize} 
\item {\bf ibrav} (integer) - {\it Default}: {\tt None} - The Bravais lattice identifier, as specified by Quantum ESPRESSO.
\item {\bf band\_path} (string) - {\it Default}: {\tt None} - String of high symmetry point labels separated by '{\tt -}' for a line connecting two points or '{\tt |}' to place the points directly adjacent on the path (see Listing 2). If {\bf band\_path} is {\tt None} the standard \AFLOW\ path will be used \cite{toher2017aflow}.
\item {\bf high\_sym\_points} (dictionary) - {\it Default}: {\tt None} - A dictionary mapping the string label of a high symmetry point to its three-dimensional crystal coordinate. (Listing 2)
\item {\bf fname} (string) - {\it Default}: {\tt 'bands'} - File name prefix for the bands. One file is written for each spin component (see listing 2).
\item {\bf nk} (int) - {\it Default}: {\tt 500} - Number of points to compute along the band path.
\end{itemize}

\subsection{\tt interpolated\_hamiltonian}
Fourier interpolation of the PAO Hamiltonian can increase the {\bf k}-grid to an arbitrary density, as described and illustrated in the manuscript for \PAOFLOW\ 1.0. The new desired dimension should be specified for {\tt nk1}, {\tt nk2}, and {\tt nk3}. The default behaviour is to double the original {\bf nk} dimension of any unspecified nfft argument. This routine populates the {\tt DataController} with a new array '{\tt Hksp}', the interpolated Hamiltonian.\\\\
Arguments for {\tt interpolated\_hamiltonian}:
\begin{itemize}
\item {\bf nfft1} (integer) - {\it Default}: {\tt None} - The desired new dimension for the Hamiltonian's previous dimension nk1. The nfft dimension should be greater than or equal to the previous nk dimension. If no argument is provided the original {\bf k}-grid dimension is doubled.
\item {\bf nfft2} (integer) - {\it Default}: {\tt None} - New interpolated dimension for nk2, following the same scheme as {\bf nfft1}.
\item {\bf nfft3} (integer) - {\it Default}: {\tt None} - New interpolated dimension for nk3, following the same scheme as {\bf nfft1} and {\bf nfft2}.
\item {\bf reshift\_Ef} (bool) - {\it Default}: {\tt False} - Shift the Hamiltonian's diagonal elements such that zero lies at the recomputed Fermi energy or at the top of the valence band.
\end{itemize}

\subsection{\tt spin\_operator}
The spin operator plays important role in several calculations performed by the\PAOFLOW\ code. Generally, when the spin operator ${\bf S}_j$ is required, \PAOFLOW\ automatically constructs it. However, ${\bf S}_j$ can be explicitly computed by calling this routine. The shell levels and their occupations are automatically read from pseudopotentials in the {\tt .save} directory.\\\\
Arguments for {\tt spin\_operator}:
\begin{itemize}
\item {\bf spin\_orbit} (bool) - {\it Default}: {\tt False} - Set this flag to {\tt True} if spin orbit coupling is added at the PAO level (with the {\tt adhoc\_spin\_orbit} routine).
\end{itemize}

\subsection{\tt add\_external\_fields}
\PAOFLOW\ supports the addition of electric fields, onsite Zeeman fields, or Hubbard corrections directly to the PAO Hamiltonian \cite{Graf:1995tl, Agapito:2015iz}. Fields must be added after the Hamiltonian's construction. Listing 3 provides an example where an electric field and Hubbard correction are simultaneously added to a Hamiltonian.\\
\label{extfield}
\begin{sexylisting}{{\tt main.py} - External fields}
from PAOFLOW import PAOFLOW

pao = PAOFLOW.PAOFLOW(savedir='system.save')
pao.projections()
pao.projectability()
pao.pao_hamiltonian()

hubbardU = np.zeros(32, dtype=float)
hubbardU[1:4] = .1
hubbardU[17:20] = 2.31
pao.add_external_fields(Efield=[.1,0.,0.], 
                        HubbardU=hubbardU)
...
\end{sexylisting}\\
Arguments for {\tt add\_external\_fields}:
\begin{itemize} 
\item {\bf Efield} (ndarray or list) - {\it Default}: {\tt [0.]} - An array of the form [$E_x$,$E_y$,$E_z$], added to the diagonal elements of the Hamiltonian. Listing 3 provides an example of adding an electric field with one non-zero component.
\item {\bf Bfield} (ndarray or list) - {\it Default}: {\tt [0.]} - An array of the form [$B_x$,$B_y$,$B_z$], specifying the strength and direction of an on-site magnetic field.
\item {\bf HubbardU} (ndarray or list) - {\it Default}: {\tt [0.]} - An array with one U entry for each orbital, e.g. [$U_1$,$U_2$,...,$U_n$] where $n$ is the number of orbitals. An example is provided in Listing 3.
\end{itemize}

\subsection{\tt adhoc\_spin\_orbit}
This routine allows the addition of spin-orbit coupling (SOC) at the PAO level. SOC is implemented for the following shell configurations, provided as the {\bf orb\_pseudo} argument: s, sp, spd, ps, ssp, sspd, \& ssppd.\\\\
Arguments for {\tt adhoc\_spin\_orbit}:
\begin{itemize}
\item {\bf naw} (ndarray or list) - {\it Default}: {\tt [1]} - List containing the number of wave functions for each pseudopotential.
\item {\bf phi} (flaot) - {\it Default}: {\tt 0.} - Spin orbit azimuthal angle.
\item {\bf theta} (float) - {\it Default}: {\tt 0.} - Spin orbit polar angle.
\item {\bf lambda\_p} (ndarray or list) - {\it Default}: {\tt [0.]} - Array of {\tt p}-orbital coupling strengths.
\item {\bf lambda\_d} (ndarray or list) - {\it Default}: {\tt [0.]} - Array of {\tt d}-orbital coupling strengths.
\item {\bf orb\_pseudo} (list) - {\it Default}: {\tt ['s']} - List of strings, containing the orbital configuration for each pseudopotential.
\end{itemize}

\subsection{\tt doubling\_Hamiltonian}
Doubles the real space dimensions of the Hamiltonian, creating a supercell in any desired direction. Naturally, the number of wavefunctions in the Hamiltonian increases by a factor $2^{\bf nx}\times2^{\bf ny}\times2^{\bf nz}$. Doubling is performed one time in each dimension by default, and doubling can be suppressed by setting the argument for a dimension to 0. \\\\
Arguments for {\tt doubling\_Hamiltonian}:
\begin{itemize}
\item {\bf nx} (int) - {\it Default}: {\tt 1} - Number of times to double the {\tt x} dimension. If {\bf nx} is set to 2 the resulting cell is 4 times larger in the {\tt x} direction.
\item {\bf ny} (int) - {\it Default}: {\tt 1} - Number of times to double the {\tt y} dimension.
\item {\bf nz} (int) - {\it Default}: {\tt 1} - Number of times to double the {\tt z} dimension.
\end{itemize}

\subsection{\tt topology}
The {\tt topology} routine calculates various quantities along the \AFLOW\ standard {\bf k}-path. If the user generates a custom path with {\tt bands}, prior to this routine, this path will be used when calculating the Z2 invariance and topological properties.

Arguments for {\tt topology}:
\begin{itemize}
\item {\bf eff\_mass} (bool) - {\it Default}: {\tt False} - Setting this flag {\tt True} computes the Hamiltonian's second derivative along the {\bf k}-path. The effective mass is calculated and saved to file with naming convention {\tt effmass\_IJ\_S.dat}. The indices {\tt I} and {\tt J} are specified by the arguments {\bf ipol} and {\bf jpol}. Index {\tt S} is specified by spin polarization at the DFT level.
\item {\bf Berry} (bool) - {\it Default}: {\tt False} - Set {\tt True} to calculate the Berry curvature along the {\bf k}-path and writes results to file as {\tt Omega\_IJ.dat}. Here, {\tt I} and {\tt J} are indices of the Berry curvature ($\Omega_{ij}$) and are specified by {\bf ipol} and {\bf jpol} respectively.
\item {\bf spin\_Hall} (bool) - {\it Default}: {\tt False} - Setting {\tt True} calculates the spin Berry curvature ($\Omega^s_{ij}$) along the {\bf k}-path and writes the results to files {\tt Omegaj\_S\_IJ.dat}. Here, the indices {\tt I}, {\tt J}, and {\tt S} are specified by the arguments {\bf ipol}, {\bf jpol}, and {\bf spol}. This routine automatically computes the Berry curvature, but no files for {\bf Berry} are written unless its flag is explicitly set {\tt True}.
\item {\bf spol} (integer) - {\it Default}: {\tt None} - Spin polarization index of the spin Berry curvature calculation. This selects which component of the spin operator is used.
\item {\bf ipol} (integer) - {\it Default}: {\tt None} - The index $i$ in effective mass and (spin) Berry calculations.
\item {\bf jpol} (integer) - {\it Default}: {\tt None} - The index $j$ in effective mass and (spin) Berry calculations.
\end{itemize}

\subsection{\tt pao\_eigh}
The routine {\tt pao\_eigh} computes the eigenspectrum for the entire {\bf k}-grid, saving the eigen-values and -vectors as new arrays in the {\tt DataController} under keys '{\tt E\_k}' and '{\tt v\_k}' respectively. Some of the previously described functions compute the eigenvalues and eigenvectors along a path, such as {\tt bands} and {\tt topology}. This routine replaces values computed by such routines with a new set of eigenfunctions, running across the entire Brillouin zone. 

No arguments are accepted when calling {\tt pao\_eigh}.

\subsection{\tt trim\_non\_projectable\_bands}
Removes eigenvalues and momenta which do not meet the projectability requirement set by {\tt projectability} from respective data arrays. This routine should be called after {\tt pao\_eigh}, if such trimming is desired.

No arguments are accepted by\\ {\tt trim\_non\_projectable\_bands}.

\subsection{\tt fermi\_surface}
Computes the bands with energies between {\bf fermi\_up} and {\bf fermi\_dw}. The results are saved in the NumPy npz format with naming convention {\tt Fermi\_surf\_band\_N\_M.npz}, where {\tt N} is the band index and {\tt M} is the spin index. The Fermi surface is saved with resolution of the existing {\bf k}-grid.\\\\
Arguments for {\tt fermi\_surface}:
\begin{itemize}
\item {\bf fermi\_up} (float) - {\it Default}: {\tt 1} - The upper energy bound for selecting bands. Bands within the range [{\bf fermi\_dw}, {\bf fermi\_up}] are included.
\item {\bf fermi\_dw} (float) - {\it Default}: {\tt 1} - The lower energy bound for selecting bands.
\end{itemize}

\begin{figure}[h!]
\begin{center}
    \includegraphics[width=0.9\columnwidth]{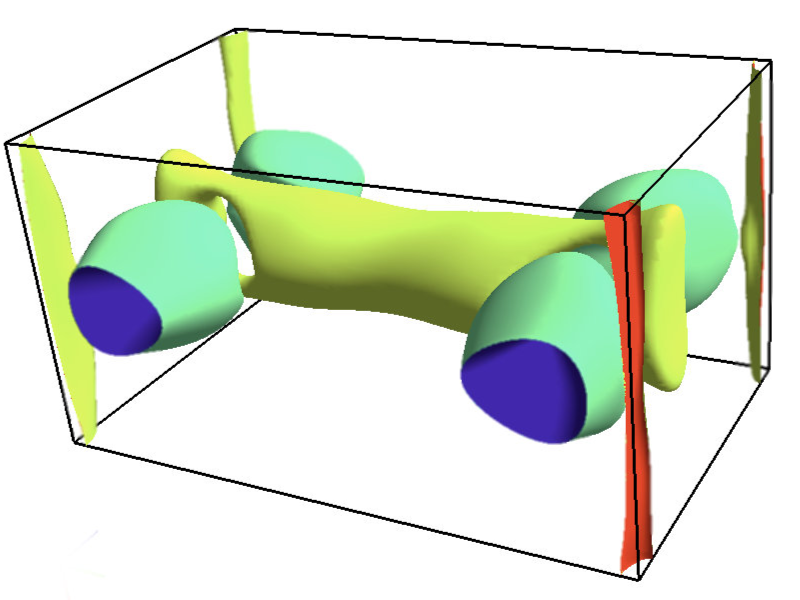}
    \vspace{-2mm}
    \caption{\small Fermi surface of FeP calculated on a ultra-dense {\bf k}-grid in PAOFLOW and visualized in FermiSurfer \cite{Kawamura2019}. For description of DFT calculations see Ref. \cite{Campbell2021}.}
    \label{fig:fermi}
  \end{center}
\end{figure}

\subsection{\tt gradient\_and\_momenta}
\label{gradient_and_momenta}
The Hamiltonian's gradient is initially calculated in real space, as it takes a simpler form. Afterward, it is Fourier transformed back into reciprocal space. Thus, the Hamiltonian's {\bf k}-space gradient is given by 

\begin{equation} \label{Hamiltonian.gradient}
\V{\nabla}_{{\bf k}} \hat{H}({\bf k}) =
\sum_{\alpha} i{\bf R} \exp\left({i{\bf k}\cdot {\bf R}}\right) \hat{H}\left({\bf R}\right)
\end{equation}
where $\hat{H}\left({\bf R}\right)$ is the real space PAO matrix and  $\left | \psi_n ({\bf k}) \right > = \exp({-i {\bf k} \cdot {\bf r}}) \left | u_n ({\bf k}) \right >$ are Bloch's functions \cite{pino}. The Hamiltonian's derivative is saved as a new array key '{\tt dHksp}' in the {\tt DataController}. 

Next, the momenta are computed from the Hamiltonian's gradient as
\begin{eqnarray}\label{momentum}
{\bf p}_{nm} ({\bf k})&=&  \left < \psi_n ({\bf k}) \right |\hat{p}\left | \psi_m ({\bf k}) \right > = \\ \nonumber
&=&  \left < u_n ({\bf k}) \right |\frac{m_0}{\hbar} \V{\nabla}_{\bf k} \hat{H}({\bf k})\left | u_m ({\bf k}) \right >
\end{eqnarray}

Additionally, the Hamiltonian's second derivative can be computed by setting the {\bf band\_curvature} argument to {\tt True}.

Arguments for {\tt gradient\_and\_momenta}:
\begin{itemize}
\item {\bf band\_curvature} (bool) - {\it Default}: {\tt False} - Compute the Hamiltonian's second derivative, stored as an array in the {\tt DataController} under the key '{\tt d2Ed2k}'.
\end{itemize}

\subsection{\tt adaptive\_smearing}
Generates the adaptive smearing parameters, stored in the {\tt DataController} as '{\tt deltakp}', used to compute quantities on energy intervals, such as the density of states or spin Hall conductivity. Allowed adaptive smearing types are gaussian ('gauss'), Methfessel-Paxton ('m-p'), or None. Implementation of the Gaussian broadening parameters are described in Section 3 of Ref \cite{PAOFLOW:2018}.\\\\
Arguments for {\tt adaptive\_smearing}:
\begin{itemize}
\item {\bf smearing} (string) - {\it Default}: 'gauss' - Method of broadening used to smooth the discrete sampling of quantities computed on energy intervals.
\end{itemize}

\subsection{\tt dos}
Compute the density of states (DOS) and/or projected density of states (PDOS) within a user defined energy range. If this routine is called after {\tt adaptive\_smearing}, the '{\tt deltakp}' smearing parameter is used to smooth the DOS calculations.\\\\
Arguments for {\tt dos}:
\begin{itemize}
\item {\bf do\_dos} (bool) - {\it Default}: {\tt True} - Flag to control whether the DOS is computed.
\item {\bf do\_pdos} (bool) - {\it Default}: {\tt True} - Flag to control whether the PDOS is computed.
\item {\bf delta} (float) - {\it Default}: {\tt 0.01} - Width of the gaussian at each energy, used to smooth the dos curves. If it has been computed with the {\tt adaptive\_smearing} routine, '{\tt deltakp}' replaces this quantity.
\item {\bf emin} (float) - {\it Default}: {\tt -10} - Lower limit for the energy range considered.
\item {\bf emax} (float) - {\it Default}: {\tt 2} - Upper limit for the energy range considered.
\item {\bf ne} (int) - {\it Default}: {\tt 1000} - The number of points to evaluate within the energy range [emin,emax].
\end{itemize}

\subsection{\tt z2\_pack}
 Writes the real space Hamiltonian to a {\tt dat} file, for use with the Z2Pack code  \cite{PhysRevB.95.075146}.\\\\
Arguments for {\tt z2\_pack}:
\begin{itemize}
\item {\bf fname} (string) - {\it Default}:\\ '{\tt z2pack\_hamiltonian.dat}' - Name for the {\tt dat} file, written to \PAOFLOW's output directory.
\end{itemize}

\begin{figure}[h!]
\begin{center}
    \includegraphics[width=0.95\columnwidth]{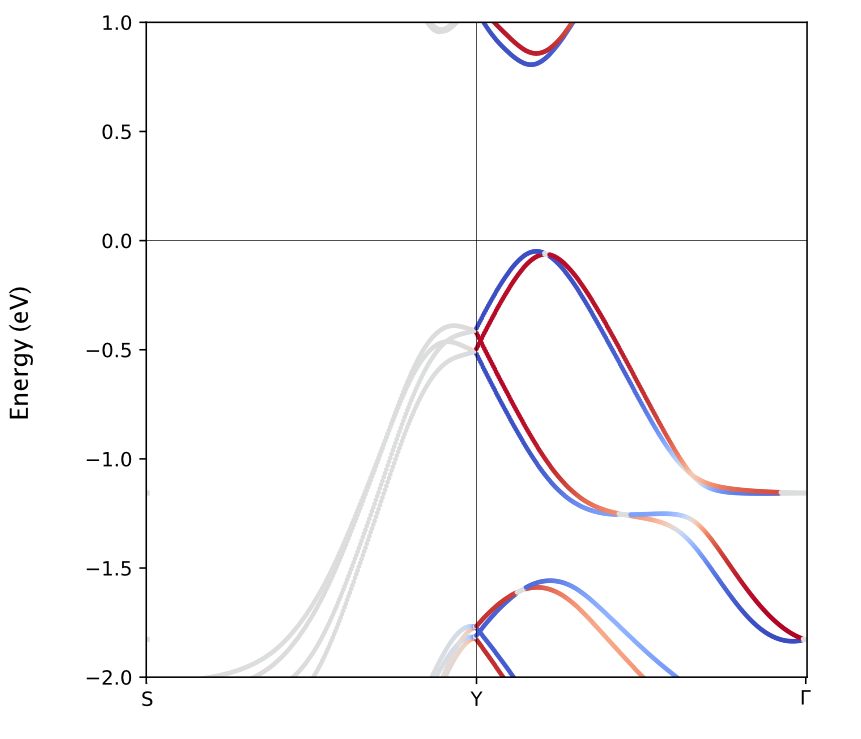}
    \vspace{-2mm}
    \caption{\small Spin texture ($S_z$) of two-dimensional ferroelectric SnTe along high-symmetry lines ({\tt example08}). See Ref. \cite{Slawinska_2020}. for details.}
    \label{fig:texture}
  \end{center}
\end{figure}

\subsection{\tt spin\_texture}
 Compute spin texture as the spin operator's expectation value for each band and for each {\bf k}-point:
\begin{equation} 
\label{spin_texture}
{\bf \Omega}_n({\bf k})=\langle \psi_n({\bf k}) | {\bf S}_j | \psi_n({\bf k}) \rangle
\end{equation}
${\bf S}_j$ is the spin operator, and $| \psi_n({\bf k}) \rangle$ are the momentum space PAO wavefunctions for band index $n$.
The spin texture is computed for bands which have values within energy range specified by {\bf fermi\_up} and {\bf fermi\_dw}. Calling this routine after {\tt bands} results in the spin texture along the same {\bf k}-path as bands, while calling it after {\tt pao\_eigh} computes spin texture across the entire BZ. Results are written to a file named {\tt spin-texture-bands.dat} for bands along a path and to {\tt NumPy .npz} format with naming convention {\tt spin\_text\_band\_N.npz} for bands across the BZ. Here, {\tt N} is the band index, and each file contains spin texture computed on \PAOFLOW's {\bf k}-grid.\\\\
Arguments for {\tt spin\_texture}:
\begin{itemize}
\item {\bf fermi\_up} (float) - {\it Default}: {\tt 1} -The spin texture is computed only for bands which contain energies beneath this upper bound.
\item {\bf fermi\_dw} (float) - {\it Default}: {\tt 1} - The spin texture is computed only for bands which contain energies above this lower bound.
\end{itemize}

\subsection{\tt anomalous\_Hall}
Calculating anomalous Hall conductivity (AHC, $\sigma_{ij}$) relies on accurate evaluation of the Berry curvature and requires a preliminary run of {\tt gradient\_and\_momenta}. \PAOFLOW\ implements a standard Kubo formula for evaluating the {\bf k}-resolved Berry curvature \cite{Yao:2004jb}. Details are given in the previous paper and in the comprehensive reference by Gradhand {\it et al.} \cite{Gradhand:2012js}. The AHC is computed with adaptive smearing, provided that the broadening parameters are calculated beforehand by {\tt adaptive\_smearing}.\\\\
Arguments for {\tt anomalous\_Hall}:
\begin{itemize}
\item {\bf do\_ac} (bool) - {\it Default}: {\tt False} - Compute the magnetic circular dichroism (MCD) on the energy range [{\bf emin}, {\bf emax}].
\item {\bf emin} (float) - {\it Default}: {\tt -1} - The minimum energy in the range on which the AHC is computed.
\item {\bf emax} (float) - {\it Default}: {\tt 1} - The maximum energy in the range. 500 points are evaluated within the interval [{\bf emin}, {\bf emax}].
\item {\bf fermi\_up} (float) - {\it Default}: {\tt 1} - Selects the upper energy bound for evaluating the Berry curvature.
\item {\bf fermi\_dw} (float) - {\it Default}: {\tt -1} - Selects the lower energy bound for evaluating the Berry curvature.
\item {\bf a\_tensor} (list) - {\it Default}: {\tt None} - List of tensor elements to evalutate. For example, setting this argument to {\tt [[0,0], [1,2]]} calculates the two components $\sigma_{xx}$ and $\sigma_{yz}$. All 9 components are computed, if the argument is left as {\tt None}.
\end{itemize}

\subsection{\tt spin\_Hall}
The spin Hall conductivity (SHC, $\sigma^{k}_{ij}$) for spin polarization along {\bf k} and charge (spin) current along i (j), is computed in a similar manner to AHC \cite{Guo:2008bx}. Here, evaluation of the {\it spin} Berry curvature is performed with the spin operator and Hamiltonian gradient as ingredients. Again, the {\tt gradient\_and\_momenta} routine is a prerequisite, and running {\tt adaptive\_smearing} beforehand controls the inclusion of broadening parameters in the calculation.

Arguments for {\tt spin\_Hall}:
\begin{itemize}
\item {\bf twoD} (bool) - {\it Default}: {\tt False} - Setting this flag {\tt True} outputs the {\tt spin\_Hall} quantities in 2-dimensional units $\Omega^{-1}$, removing any dependence on the sample height. It is assumed that the dimensions of interest are oriented in the {\tt x}-{\tt y} plane, and the slab is oriented along {\tt z}.
\item {\bf do\_ac} (bool) - {\it Default}: {\tt False} - Compute the spin circular dichroism (SCD) on the energy range [{\bf emin}, {\bf emax}].
\item {\bf emin} (float) - {\it Default}: {\tt -1} - The minimum energy in the range on which the SHC and SCD are computed.
\item {\bf emax} (float) - {\it Default}: {\tt 1} - The maximum energy in the range. Again, 500 points are evaluated in the interval [{\bf emin}, {\bf emax}].
\item {\bf fermi\_up} (float) - {\it Default}: {\tt 1} - Selects the upper energy bound for evaluating the spin Berry curvature.
\item {\bf fermi\_dw} (float) - {\it Default}: {\tt -1} - Selects the lower energy bound for evaluating the spin Berry curvature.
\item {\bf s\_tensor} (list) - {\it Default}: {\tt None} - List of tensor elements to evalutate. To calculate $\sigma^{x}_{xx}$ and $\sigma^{z}_{xy}$ components use {\tt[[0,0,0],[0,1,2]]}. If the argument is left as {\tt None}, all 27 components are computed.
\end{itemize}

\subsection{\tt doping}
Determine the chemical potential corresponding to a specified doping concentration and temperature range.\\\\
Arguments for {\tt doping}:
\begin{itemize}
\item {\bf tmin} (float) - {\it Default}: {\tt 300} - Minimum temperature for which to evaluate the chemical potential.
\item {\bf tmax} (float) - {\it Default}: {\tt 300} - Maximum temperature to compute chemical potential.
\item {\bf nt} (int) - {\it Default}: {\tt 1} - The number of temperatures to evaluate in the range [{\bf tmin}, {\bf tmax}].
\item {\bf delta} (float) - {\it Default}: {\tt 0.01} - Gaussian broadening width, used to smooth the density of states along the energy range. Doping calculation involves an integration over density of states and therefore includes a call to the dos module.
\item {\bf emin} (float) - {\it Default}: {\tt -1} - Lowest value of energy of the occupied bands.
\item {\bf emax} (float) - {\it Default}: {\tt 1} - At least the energy of the minimum of the conduction bands to obtain accurate results.
\item {\bf ne} (int) - {\it Default}: {\tt 1000} - Number of points in the energy grid.
\item {\bf doping\_conc} (float) - {\it Default}: {\tt 0} - The doping concentration in carriers/cm${^3}$ for which to compute the chemical potential. Specify negative value for n-type doping and positive value for p-type doping.
\item {\bf core\_electrons} (int) - {\it Default}: {\tt 0} - If the total number of electrons in the lower energy bands is known, this value can be introduced here. In this case, {\bf emin} does not have to be the lowest energy value of occupied bands but can instead be set above energies of the core bands, to speed up integration.
\item {\bf fname} (string) - {\it Default}: '{\tt doping\_}' - Prefix for the output file containing doping versus temperature.
\end{itemize}
 
\subsection{\tt density}
\begin{figure}[h!]
\begin{center}
    \includegraphics[width=0.95\columnwidth]{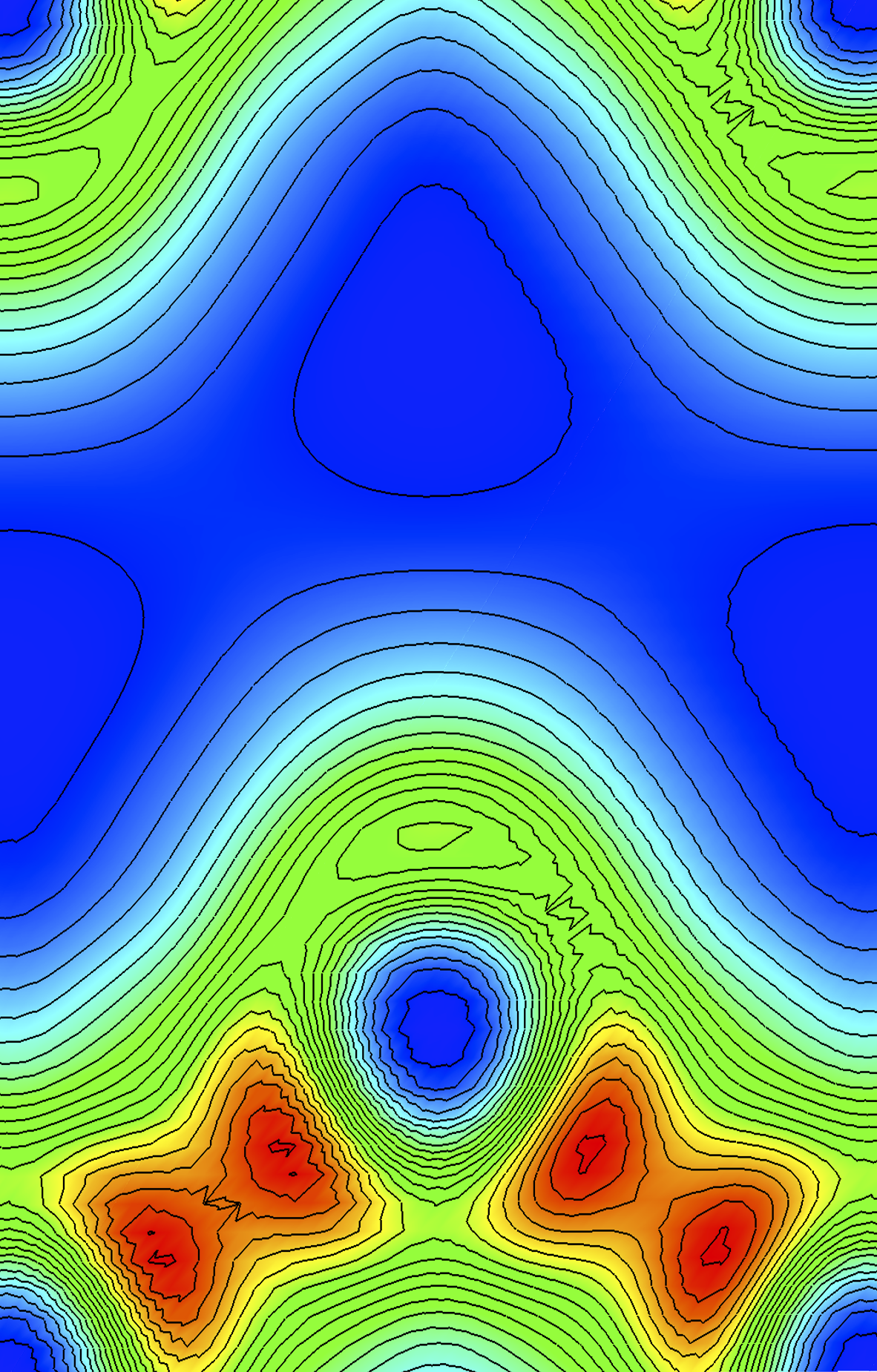}
    \vspace{-2mm}
    \caption{\small Electronic density for diamond structure of silicon on the $\langle 1, 0, -1\rangle$ plane cut, calculated from the real space PAO wavefunctions ({\tt example01}).}
    \label{fig:silicon_electronic_density}
  \end{center}
\end{figure}

Calculate the electronic density on a real space grid, performed for silicon in Listing 4 and displayed in Figure \ref{fig:silicon_electronic_density}. Wavefunctions in {\bf k}-space (produced by {\tt pao\_eigh}) are required as a prerequisite, and the PAO projections {\it must} be performed by \PAOFLOW's {\tt projections} method. This algorithm serves as a recipe for constructing the real space PAO wavefunctions. Although this is currently the only routine to utilize such construction, future versions of \PAOFLOW\ will include other methods for computing spatially resolved quantities. The real space grid dimension defaults to 48x48x48 but can be specified with the optional arguments {\bf nr1}, {\bf nr2}, and {\bf nr3}.

Arguments for {\tt density}:
\begin{itemize}
\item {\bf nr1} (int) - {\it Default}: {\tt 48} - Number of points in the first dimension of the real space grid, over which to compute the charge density.
\item {\bf nr2} (int) - {\it Default}: {\tt 48} - Number of points in the second dimension of the real space grid.
\item {\bf nr3} (int) - {\it Default}: {\tt 48} - Number of points in the third dimension of the real space grid.
\end{itemize}

\begin{sexylisting}{{\tt main.py} - Density}
from PAOFLOW import PAOFLOW

pao = PAOFLOW.PAOFLOW(savedir='system.save')
pao.projections()
pao.projectability()
pao.pao_hamiltonian()
pao.pao_eigh()

pao.density(nr1=48, nr2=48, nr3=48)

pao.finish_execution()
\end{sexylisting}

\subsection{\tt transport}

Calculate the transport properties, such as electrical conductivity, Seebeck coefficients, and thermal conductivity. The transport properties are computed in the constant relaxation time approximation, unless built-in or user defined $\tau$ models are provided. See \verb|example09| for detailed information on specifying models for the relaxation time $\tau$.\\\\
Arguments for {\tt transport}:
\begin{itemize}
\item {\bf tmin} (float) - {\it Default}: {\tt 300} - Minimum temperature for which to evaluate transport properties.
\item {\bf tmax} (float) - {\it Default}: {\tt 300} - Maximum temperature to evaluate transport properties.
\item {\bf nt} (float) - {\it Default}: {\tt 1} - The number of temperatures to evaluate in the range [{\bf tmin},{\bf tmax}].
\item {\bf emin} (float) - {\it Default}: {\tt 1} - Minimum value in the energy grid [{\bf emin},{\bf emax}].
\item {\bf emax} (float) - {\it Default}: {\tt 10} - Maximum value of energy in the grid.
\item {\bf ne} (int) - {\it Default}: {\tt 1000} - Number of points in the energy range [{\bf emin},{\bf emax}].
\item {\bf scattering\_channels} (list) - {\it Default}: {\tt None} - List of strings and/or {\tt TauModel} objects containing the scattering models to be included in the calculation of $\tau$ .
\item {\bf scattering\_weights} (list) - {\it Default}: {\tt None} - Initial guess for the parameters $a_{imp}$, $a_{ac}$, $a_{op}$ etc to be used for the fitting procedure if fit is set to {\tt True}. The default behavior with this argument set to {\tt None} is to use unity as every scattering weight.
\item {\bf tau\_dict} (dict) - {\it Default}: {\tt \{\}} - Dictionary of parameters required for the calculation of scattering models.
\item {\bf write\_to\_file} (bool) - {\it Default}: {\tt True} - Controls the output of the several data fields produced by this routine. No files are written if the flag is set to {\tt False}.
\item {\bf save\_tensors} (bool) - {\it Default}: {\tt False} - Setting this flag {\tt True} stores the resulting electrical conductivity, Seebeck coefficient, and thermal conductivity to the data\_arrays dictionary with respective keys 'sigma', 'S', and 'kappa'. 
\end{itemize}

\subsection{\tt find\_weyl\_points}
Perform a search for Weyl points within the first Brillouin zone. The search identifies Weyl point candidates by utilizing {\tt Scipy}'s {\tt minimize} function with the 'L-BFGS-B' algorithm.\\\\
Arguments for {\tt find\_weyl\_points}:
\begin{itemize}
\item {\bf symmetrize} (bool) - {\it Default}: {\tt False} - Use QE symmetry operations to unfold equivalent {\bf k}-points. If equivalent {\bf k}-points are Weyl points, all such points are reported.
\item {\bf search\_grid} (list) - {\it Default}: {\tt [8, 8, 8]} - Dimensions of the grid on which the minimization routine is performed. Bands are Fourier interpolated on this grid to improve resolution.
\end{itemize}

\subsection{\tt restart\_dump}
\PAOFLOW's computational state can be saved at any time with the {\tt restart\_dump} routine. Data is stored in the {\tt json} format, and the naming convention for such files can be chosen with the {\bf fname\_prefix} argument. Each processor saves a file in the {\bf workpath} directory with name {\tt fname\_prefix\_N.json}, where {\tt N} is the core's rank. For this reason, restarted calculations must be executed with the same number of cores. See Listing 5 for an example where the gradient and momenta are computed and dumped for reuse in another calculation.\\\\
Arguments for {\tt restart\_dump}:
\begin{itemize}
\item {\bf fname\_prefix} (string) - {\it Default}: {\tt 'paoflow\_dump'} - The name prefix for the dump files, written to the working directory.
\end{itemize}
\begin{sexylisting}{{\tt main.py} - Restart (dump)}
from PAOFLOW import PAOFLOW

pao = PAOFLOW.PAOFLOW(savedir='system.save')
pao.projections()
pao.projectability()
pao.pao_hamiltonian()
pao.pao_eigh()
pao.gradient_and_momenta()
pao.restart_dump(fname_prefix='pao')
\end{sexylisting}

\subsection{\tt restart\_load}
Recover \PAOFLOW's calculation state from a previous run, saved to the json format with {\tt restart\_dump}. A restarted run must be executed with the same number of cores as the run which produced the dump files. Listing 6 provides example usage for {\tt restart\_load}.\\\\
Arguments for {\tt restart\_load}:
\begin{itemize}
\item {\bf fname\_prefix} (string) - {\it Default}: {\tt 'paoflow\_dump'} - The name prefix for the dump files, written to the working directory.
\end{itemize}
\begin{sexylisting}{{\tt main.py} - Restart (load)}
from PAOFLOW import PAOFLOW

pao = PAOFLOW.PAOFLOW(restart=True)
pao.restart_load(fname_prefix='pao')
pao.adaptive_smearing()
...
\end{sexylisting}

\subsection{\tt finish\_execution}
Conclude the \PAOFLOW\ run and remove references to memory intensive quantities. Details about the execution are provided, such as run duration and total memory requirements. This routine should be called once all desired calculations are performed for a given \PAOFLOW\ object, especially if the code continues to create other \PAOFLOW\ Hamiltonians.\\\\
{\tt finish\_execution} accepts no arguments.

\section{Tight-binding models}
\label{tbmodels}

\PAOFLOW\ is capable of generating a Hamiltonian from analytical TB models, such as the Kane-Mele or Slater-Koster models \cite{Slater_Koster_1954,Kane_2005}. For each type of model, one needs to specify a few system properties, such as hopping parameters, lattice constant, etc. These parameters, including the label selecting the model to implement, should be initially stored in a dictionary, which is subsequently passed into \PAOFLOW's constructor as the {\bf model} argument. Once the Hamiltonian is constructed, \PAOFLOW's class methods can be applied in the standard way to compute any desired quantities. Two examples are provided in Listing 7 and 8, and more are provided in the {\tt examples/} directory.

\subsection{Cubium}
Creates a Hamiltonian for a single atom in the simple-cubic geometry, containing one orbital per site. The hopping parameter is defined by including an entry in the parameters dictionary with key '{\tt t}', and should have units of eV.\\\\
Required dictionary entries:
\begin{itemize}
\item {\it Key}: '{\tt label}' - Keyword identifier for the model: '{\tt cubium}', in this case. The labels are not case sensitive.
\item {\it Key}: '{\tt t}' - The hopping parameter for nearest neighbor interactions, in units of eV.
\end{itemize}

\subsection{Cubium II}
Creates a Hamiltonian for a single atom in the simple-cubic geometry, implementing the double band model with two orbitals per site. The hopping parameter and band gap energy are given by '{\tt t}' and '{\tt Eg}' respectively.\\
Required dictionary entries:
\begin{itemize}
\item {\it Key}: '{\tt label}' - Keyword identifier for the model: '{\tt cubium2}'
\item {\it Key}: '{\tt t}' - The hopping parameter for nearest neighbor interactions.
\item {\it Key}: '{\tt Eg}' - Band gap energy, in eV.
\end{itemize}

\subsection{Graphene}
\label{tbm-graphene}
A simple TB model for graphene, considering only nearest neighbor interactions. The hopping parameter is specified with parameter '{\tt t}'. The lattice constant is taken as $a=2.46$~\AA, and lattice vectors are given in a standard form: ${\bf a_1}=a \langle 1, 0, 0 \rangle$, ${\bf a_2}=a \langle \frac{1}{2}, \frac{\sqrt{3}}{2}, 0 \rangle$, ${\bf a_3}=a \langle 0, 0, 10 \rangle$.\\\\
Required dictionary entries:
\begin{itemize}
\item {\it Key}: '{\tt label}' - Keyword identifier for the model: '{\tt graphene}'
\item {\it Key}: '{\tt t}' - The hopping parameter for nearest neighbor interactions.
\end{itemize}

\subsection{Kane-Mele model}
Constructs a Kane-Mele Hamiltonian for graphene. The first nearest neighbors are handled in the standard manner, with hopping parameter '{\tt t}'. Second nearest neighbors are treated with spin dependent amplitude, characterized by the parameter '{\tt soc\_par}'. See an example in Listing 7.\\\\
Required dictionary entries:
\begin{itemize}
\item {\it Key}: '{\tt label}' - Keyword identifier for the model: '{\tt kane\_mele}'.
\item {\it Key}: '{\tt alat}' - The lattice parameter, $a$. The lattice vectors are the same as in section \ref{tbm-graphene}.
\item {\it Key}: '{\tt t}' - The hopping parameter for nearest neighbor interactions.
\item {\it Key}: '{\tt soc\_par}' - The spin-orbit coupling parameter for second nearest neighbor interactions.
\end{itemize}
\begin{sexylisting}{{\tt main.py} - Kane-Mele model}
from PAOFLOW import PAOFLOW

model = {'label':'Kane_Mele', 't':1.0,
         'soc_par':0.1, 'alat':1.0}

paoflow = PAOFLOW.PAOFLOW(model=model, 
                  outputdir='./kane_mele')
...
\end{sexylisting}

\subsection{Slater-Koster model}
A generalized Slater-Koster TB model in the two-center approximation, that includes only {\tt s} and {\tt p} orbitals of first nearest neighbors. The user must specify the lattice vectors, the atomic positions, the included orbitals for each atom, and the hopping parameters. See Listing 8 for further details.\\\\
Required dictionary entries:
\begin{itemize}
\item {\it Key}: '{\tt label}' - Keyword identifier for the model: '{\tt slater\_koster}'.
\item {\it Key}: '{\tt a\_vectors}' - A numpy array containing the three primitive lattice vectors.
\item {\it Key}: '{\tt atoms}' - A dictionary with entries specifying atomic information for each atom. Dictionary keys label the atoms numerically with strings (e.g. the first atom has key '{\tt 0}'), and the corresponding values are dictionaries with information about the atom. The species, position (in crystal coordinates), and string identifier for each represented orbital should be saved in the atomic dictionary with respective keys: '{\tt name}', '{\tt tau}', and '{\tt orbitals}'. The name is simply a string, the atomic position is a 3-vector, and orbitals are a list of strings denoting the orbitals belonging to each atom. 
\item {\it Key}: '{\tt hoppings}' - A dictionary defining different hopping parameters, in eV. The Slater-Koster hopping parameters should be labeled: {\tt sss}, {\tt sps}, {\tt pps}, and {\tt ppp}.
\end{itemize}
\begin{sexylisting}{{\tt main.py} - Slater-Koster model}
from PAOFLOW import PAOFLOW
import numpy as np

model = {'label':'slater_koster'}

avecs = np.array([[.5,.5,0],
                  [.5,0,.5],
                  [0,.5,.5]])

atoms = {'0' : 
         {'name':'Si',
          'tau':[0,0,0],
          'orbitals':['s','px','py','pz']},
         '1' :
         {'name':'Si',
          'tau':[.25,.25,.25],
          'orbitals':['s','px','py','pz']}}

hops = {'sss':-2.36233, 'sps': 1.86401,
        'pps': 2.85882, 'ppp':-0.94687}

model['a_vectors'] = avecs
model['atoms'] = atoms
model['hoppings'] = hops

paoflow = PAOFLOW.PAOFLOW(model=model, 
                  outputdir='./kane_mele')
...
\end{sexylisting}

\section{Scattering models}
\PAOFLOW\ supports a diverse set of scattering effects by allowing users to implement temperature and energy dependent models for the relaxation time parameter $\tau$. Functional models are defined with the {\tt TauModel} class. There are many built-in models, which only require the specification of empirical constants, and users can define new models to pass into \PAOFLOW\ directly. A Python dictionary containing required parameters for any selected built-in models must be passed to the {\it transport} routine as the {\bf tau\_dict} argument (see Listing 9). Table 1 details the various constant parameters and their key strings for dictionary entries. Models for $\tau$ models are necessarily dependent on two quantities, the temperature and the Hamiltonian's energy eigenvalues. Other varying parameters can be supplied to {\tt TauModel}s through the params dictionary. As such, a python function accepting three arguments (the temperature, the energy, and the parameters dictionary) is the required format when constructing a custom {\tt TauModel} object. The $\tau$ for each included model are computed, by evaluating the {\tt TauModel} functions. Then, the $\tau$s are harmonically summed to obtain the effective $\tau$ for all scattering channels. The functional form for a {\tt TauModel} is presented in Listing 8 and a usage case in \verb|example10|.

\begin{sexylisting}{Define TauModel}
from PAOFLOW.defs.TauModel import TauModel
from PAOFLOW import PAOFLOW

# Compute quantites required for transport
pao = PAOFLOW.PAOLFOW(savedir='system.save')
pao.projections()
pao.projectability()
pao.pao_hamiltonian()
pao.pao_eigh()
pao.gradient_and_momenta()
pao.adaptive_smearing()

# Define the functional model
#  for acoustic scattering.
rho = 5.3e3; v = 5.2e3; D = 7*1.6e-19
m = .7*9.11e-13; h_bar = 6.58e-16;
ac_const = 2*np.pi*h_bar**4*rho*v**2
ac_const /= ((2*m)**(3/2)*D**2)
def acoustic_scat ( temp, ene, params ):
  return ac_const/(temp*np.sqrt(ene))

# Define TauModel object
ac_model = TauModel(function=acoustic_scat)

channels = [ac_model, 'optical']

# Define parameters for built-in models
tau_params = {'ms':0.7', 'hwlo':[0.03536]
              'eps_inf':11.6, 'eps_0':13.5}

pao.transport(scattering_channels=channels,
              tau_dict=tau_params)

pao.finish_execution()
\end{sexylisting}\\

\subsection{Charged impurity scattering}
In order to include the effect of electron scattering from impurities, include '{\tt impurity}' in the list  {\bf scattering\_channels} \cite{jacoboni:2010theory,Fiorentini:96}. This calculates the relaxation time as
\begin{equation} 
\label{Impurity scattering}
\tau_{im}(E,T) = \frac{E^{\frac{3}{2}} \sqrt{2m^{*}}4\pi\varepsilon^2} {(log(1+\frac{1}{x})-\frac{1}{1+x})\pi n_I Z_I^{2}e^{4}}\\
\end{equation}

\begin{equation} 
\label{Impurity scattering x} 
x = \frac{E}{k_{B}T} \\
\end{equation}
Required parameters are $m^{*}$, $\varepsilon_{0}$, $\varepsilon_{\inf}$, $n_{I}$, $Z_{I}$.

\subsection{Acoustic scattering}
In order to include the effect of electron scattering from acoustic phonons, include '{\tt acoustic}' in the list  {\bf scattering\_channels}. This calculates the relaxation time according to \cite{jacoboni:2010theory,Fiorentini:96}
\begin{equation} 
\label{Acoustic scattering}
\tau_{ac}(E,T)= \frac{2\pi\hbar^{4}\rho v^{2} }{(2m^{*})^{\frac{3}{2}}k_BT D_{ac}^{2} \sqrt E} \\
\end{equation}
Required parameters are $m^{*}, \rho, v, D_{ac}$.

\subsection{Optical scattering}
In order to include the effect of electron scattering from optical phonons, include '{\tt optical}' in the list  {\bf scattering\_channels}. This calculates the relaxation time following \cite{jacoboni:2010theory,Fiorentini:96}
\begin{equation} 
\label{Optical scattering}
\tau_{op}(E,T)= \frac{\sqrt{2k_{B}T}\pi x_{o}\hbar^{2}\rho}{{m^{*}}^{\frac{3}{2}}{D_{op}}^{2}N_{op}\sqrt{x+x_{o}}+(N_{op}+1)\sqrt{x-x_{o}}} \\
\end{equation}
\begin{equation}
    N_{op} = \frac{1}{\exp\frac{{\hbar\omega_{l}}}{k_B{T}}-1},  \\
    x = \frac{E}{k_{B}T},  \\
    x_{o} = \frac{\hbar\omega_{l}}{k_{B}T},  \\
\end{equation}
Required parameters are $m^{*}, \rho, \omega_{l}, D_{op}, N_{op}$.

\subsection{Polar acoustic scattering}
In order to include the effect of electron scattering from acoustic phonons in polar materials, include '{\tt polar\_acoustic}' in the list  {\bf scattering\_channels}. This calculates the relaxation time following \cite{jacoboni:2010theory,Fiorentini:96}

\begin{equation} 
\label{Polar acoustic scattering}
\begin{split}
\tau_{pac}(E,T) = \frac{\sqrt{2E}2\pi\varepsilon^2\hbar^2\rho v^2}{p^2 e^2 \sqrt{m^*} k_BT} \\
 \times \left[1-\frac{\epsilon_{o}}{2E}\log(1+4\frac{E}{\epsilon_{o}})+\frac{1}{1+4\frac{E}{\epsilon_{o}}}\right]
\end{split}
\end{equation}
Required parameters are $m^{*}, \varepsilon_{0}, \varepsilon_{\inf}, \rho, v, p$. 

\subsection{Polar optical scattering}
In order to include the effect of electron scattering from optical phonons in polar materials, include '{\tt polar\_optical}' in the list  {\bf scattering\_channels}. This calculates the relaxation time based on \cite{jacoboni:2010theory,Fiorentini:96,ridley:1998polar}.

\begin{equation} 
\label{Polar optical scattering}
\begin{split}
\tau_{pop}(E,T) = \\
 \sum_{i}\frac{Z(E,T,\omega_{i}^{l})E^{\frac{3}{2}}}{C(E,T,\omega_{i}^{l})-A(E,T,\omega_{i}^{l})-B(E,T,\omega_{i}^{l})}
\end{split}
\end{equation}

\begin{equation} 
\begin{aligned}
\label{Polar optical scattering A}
A(E,T,\omega_{l}) =n(\omega_{l}+1)\frac{f_{0}(E+\hbar\omega_{l})}{f_{0}(E)}[(2E+\hbar\omega_{l})\\
sinh^{-1}(\frac{E}{\hbar\omega_{l}})^{\frac{1}{2}}-[E(E+\hbar\omega_{l})]^\frac{1}{2}]
\end{aligned}
\end{equation}
\begin{equation} 
\begin{aligned}
\label{Polar optical scattering B}
B(E,T,\omega_{l}) =\theta(E-\hbar\omega_{l})n(\omega_{l})\frac{f_{0}(E-\hbar\omega_{l})}{f_{0}(E)}\\
[(2E-\hbar\omega_{l})cosh^{-1}(\frac{E}{\hbar\omega_{l}})^{\frac{1}{2}}\\
-[E(E-\hbar\omega_{l})]^\frac{1}{2}]
\end{aligned}
\end{equation}
\begin{equation} 
\begin{aligned}
\label{Polar optical scattering C}
C(E,T,\omega_{l}) =2E[n(\omega_{l}+1)\frac{f_{0}(E+\hbar\omega_{l})}{f_{0}(E)}\\
sinh^{-1}(\frac{E}{\hbar\omega_{l}})^{\frac{1}{2}}+\theta(E-\hbar\omega_{l})n(\omega_{l})\\
\frac{f_{0}(E-\hbar\omega_{l})}{f_{0}(E)}cosh^{-1}(\frac{E}{\hbar\omega_{l}})^{\frac{1}{2}}]
\end{aligned}
\end{equation}
\begin{equation} 
\begin{aligned}
\label{Polar optical scattering Z}
Z(\omega_{l}) =\frac{2}{W_{0}(\hbar \omega_{l})^\frac{1}{2}},
\,W_{0}(\omega_{l})=\frac{e^{2}\sqrt{2m^{*}\omega_{l}}\varepsilon^{-1}}{4\pi\hbar^{\frac{3}{2}}}
\end{aligned}
\end{equation}

Required parameters are $m^{*}, \varepsilon_{0}, \varepsilon_{\inf}, \omega_{LO}$. 

\begin{table}[h!]
  \begin{tabular}{|p{0.22\textwidth}ccc|}
    \hline
     Parameter & Symbol & Units & Key \\ 
     \hline
     Mass density & $\rho$ & $kg/m^{3}$ & $rho$  \\ 
     Low freq. dielectric constant & $\epsilon_{0}$ & - & eps\_0 \\
     High freq. dielectric constant & $\epsilon_{\infty}$ & - & eps\_inf \\
     Acoustic velocity & $v$ & m/s & v\\
     Effective mass ratio & $m^{*}$ & - & ms \\
     Acoustic deformation potential & $D_{ac}$ & eV & D\_ac \\
     Optical deformation potential & $D_{op}$ & eV & D\_op \\
     Optical phonon energy & $\hbar\omega_{l}$ & eV & hwlo \\
     Number of impurities & $n_{I}$ & $cm^{-3}$ & nI \\
     Charge on impurity & $Z_{I}$ & - & Zi \\
     Piezoelectric constant & $p$ & $C/m^{2}$ & piezo \\
    \hline
  \end{tabular}
  \caption{Symbols, units, and corresponding key in {\bf tau\_dict} for the parameters required in various scattering models.}
  \label{Scattering parameters}
\end{table}

\subsection{Effective scattering time}
\begin{equation}
  \label{matths_eqn}
\frac{1}{\tau_{total}(E,T)} =\frac{1}{\tau_{im}(E,T)}+\frac{1}{\tau_{ac}(E,T)}+...
\end{equation}
The total scattering time is calculated as a harmonic sum of the specified scattering mechanisms, evaluated for each energy and temperature, Eq \ref{matths_eqn}.
The effective scattering time $\tau_{total}$ is computed for each energy and temperature during the execution of the {\tt transport} routine.
\section{Performance}
\begin{figure}[h!]
\begin{center}
    \includegraphics[width=1.0\columnwidth]{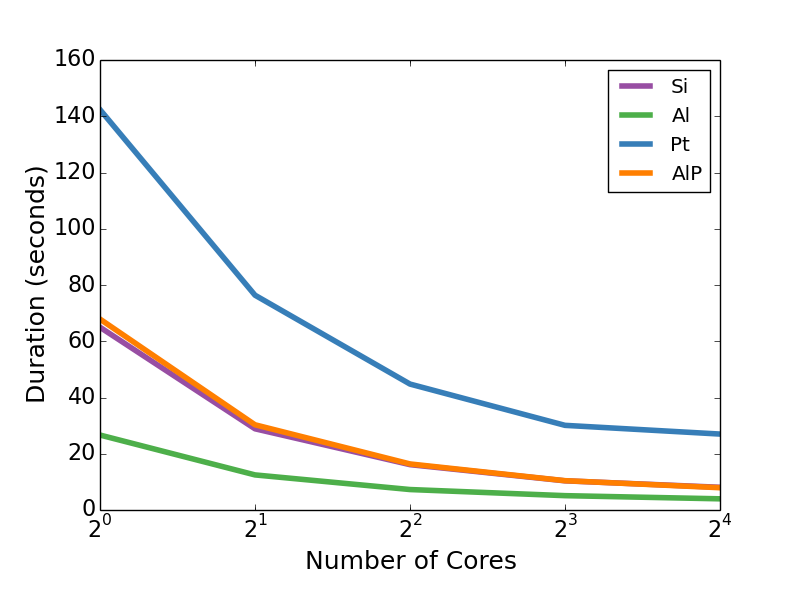}
    \vspace{-2mm}
    \caption{\small Selected examples (see {\tt examples/} on GitHub) performed on an increasing number of processors. Parallelized routines provide run time scaling nearly proportional to the number of cores used in a calculation, closely approaching the speed increase limit of Amdahl's Law.}
    \label{corepfig}
  \end{center}
\end{figure}

\begin{figure}[h!]
\begin{center}
    \includegraphics[width=1.0\columnwidth]{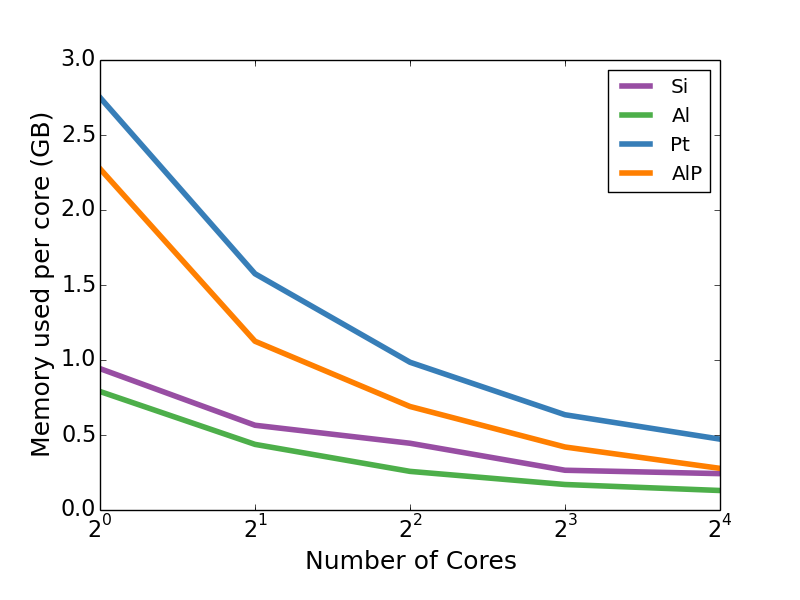}
    \vspace{-2mm}
    \caption{\small Memory scaling per core (in GB), for selected examples. An increasing core count reduces the memory requirements per processor.}
    \label{memfig}
  \end{center}
\end{figure}

Benchmark performance tests reveal excellent scaling for massively parallelized calculations. \PAOFLOW\ exploits parallelization over bands whenever possible, primarily in the calculation of gradients. However, most routines are parallelized across the {\bf k}-point mesh or path. \PAOFLOW\ also possesses excellent scaling of memory requirements, in parallel runs. Increasing the number of processors can reduce the memory load on each processor, as many of the large arrays are distributed evenly among the cores. Performance is analyzed on a Dell PowerEdge R730 server with two 2.4GHz Intel Xeon E5-2680 v4 fourteen-core processors, and results for several examples are presented in Fig \ref{corepfig} and Fig \ref{memfig}. \PAOFLOW\ demonstrates excellent scaling properties on manycore systems and possesses massively parallel capabilities.

\section{Conclusions}
\PAOFLOW\ provides a lightweight, robust tool for efficient materials and Hamiltonian analysis. Continuous development of the package has streamlined its functionality and enabled many new tools for effectively characterizing the electronic properties of solids. The updated framework offers an ideal tool for high throughput condensed matter simulations and generation for materials genomics \cite{aflowBZ}.

\section{Acknowledgments}
 We are grateful for computational resources provided by the High Performance Computing Center at the University of North Texas and the Texas Advanced Computing Center at the University of Texas, Austin.
The members of the \AFLOW\ Consortium  (http://www.aflow.org)
acknowledge support  by DOD-ONR (N00014-13-1-0635, N00014-11-1-0136,
N00014-15-1-2863). The authors also acknowledge Duke University Center for Materials Genomics.

\section{Data Availability}
Data generated for this manuscript's production is available on {\tt GitHub} (\verb|https://github.com/marcobn/PAOFLOW|), provided within the various examples directories outlined in the text above. 
\newcommand{\Ozolins}{Ozoli\c{n}\v{s}}

\end{document}